\documentclass[twocolumn,showpacs,pra]{revtex4}

\usepackage{amsfonts} 
\usepackage{amsmath}
\usepackage{amssymb}
\usepackage{txfonts}
\usepackage{graphicx}
\usepackage{graphics}
\usepackage{dcolumn}
\usepackage{bm}
\begin{document}

\preprint{APS/123-QED}

\title{Optomechanical Akhmediev Breather}
\author{Hao Xiong}\email{haoxiong@hust.edu.cn}
\author{Ying Wu}
\affiliation{School of Physics, Huazhong University of Science and Technology, Wuhan, 430074, P. R. China}
\date{\today}

\begin{abstract}
Akhmediev breather has attracted considerable attention over the past decades in various fields such as hydrodynamics, plasma physics, and nonlinear optics because its wide applications and its importance in understanding nonlinear coherent phenomena. Here we introduce optomechanical arrays as a novel and integrated platform to observe both photonic and phononic Akhmediev breather, with remarkable functionality for readout and control complementing existing research on hydrodynamic and plasma systems. Our results pave a path toward reconfigurable nano- photonic and phononic networks, and may enable a new class of devices in signal processing and on-chip nonlinear dynamical systems.

\end{abstract}

\maketitle

\begin{figure}[ht]
\centering
\includegraphics[width=0.47\textwidth]{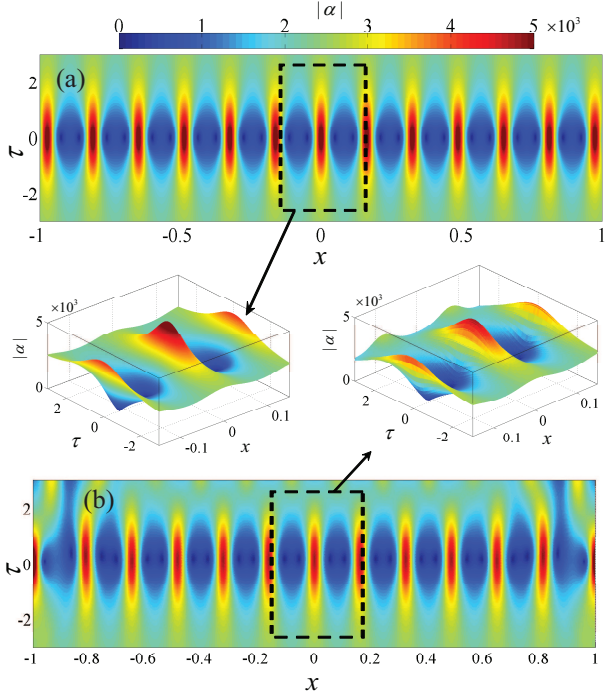}
\caption{\label{fig:1} Temporal evolution of (a) the exact Akhmediev breather solution and (b) cavity fields in the optomechanical array with $N=500$.}
\end{figure}

Akhmediev breather \cite{Akhmediev} is one of the fascinating phenomena in nonlinear science. It undergoes a single growth-return cycle in time and exhibits a periodic structure in space. Typical characteristic for an Akhmediev breather is shown in Fig. \ref{fig:1}(a), where $x$ and $\tau$ are the normalized space and time coordinate, respectively. Over the past few decades, Akhmediev breather has played a fundamental role in the development of nonlinear science and attracted significant interest in various subjects, including hydrodynamics \cite{Akhmediev-Hydrodynamics}, turbulence \cite{Akhmediev-Turbulence}, and atmosphere science \cite{Akhmediev-Atmosphere}. The paradigm is essential for understanding nonlinear localization and evolution beyond conventional linear system for infinitesimal amplitudes because it allows an analytical description of the temporal-spatial structure of the sideband modes. In addition, some studies \cite{Akhmediev-optics} have shown that Akhmediev breather also provides new insights into the development of supercontinuum generation due to the localized temporal structure.

It is well known that integrated photonics and phononics have fostered numerous chip-scale sensing and signal processing technologies \cite{om}, and on-chip devices with controllable photonic or phononic localization are vital for information processing \cite{loc}. Akhmediev breather provides an accessible route to this objective due to its excellent localized properties and tunable functionality \cite{Akhmediev-rev}. However, generation of Akhmediev breather based on conventional mechanisms, including hydrodynamics \cite{Akhmediev-Hydrodynamics} and plasma oscillation \cite{Akhmediev-Atmosphere}, can not be easily integrated with nanoscale photonic and phononic circuits. In fact, it remains an outstanding challenge to create photonic/phononic Akhmediev breather in an on-chip solid-state construction \cite{Akhmediev-rev}.

In the present work, we introduce micro-fabricated optomechanical arrays as a novel and integrated platform to observe both photonic and phononic Akhmediev breather (which is so-called optomechanical Akhmediev Breather), with remarkable functionality for readout and control complementing existing research on hydrodynamic and plasma systems. Both photonic and phononic Akhmediev breather solutions are obtained analytically in the non-perturbative regime and examined by the Lie transformation. Further more, numerical simulations with different parameters show an excellent agreement with this theory. Optomechanical Akhmediev Breather may find many applications in on-chip manipulation of mechanical oscillation and light propagation because micro-fabricated optomechanical arrays are compatible with a wide variety of on-chip photonic and phononic devices.

Optomechanical interaction \cite{rev} has attracted significant interest due to their important applications in force sensors and precision measurements \cite{om2,sensor}, and enables a new platform for achieving on-chip manipulation of light propagation \cite{omit,omit2}. Owing to the concept of optomechanical crystal \cite{crystals}, the ability to create periodic arrays of coupled optomechanical devices in a Si microchip has been demonstrated experimentally \cite{crystals2,crystals3}. Here we consider the optomechanical array which consists of $N$ optomechanical systems. A mechanical mode and an optical mode are well-coupled in each optomechanical system via resonantly enhanced feedback-backaction arising from radiation pressure, and neighboring optomechanical systems are coupled via photon hopping. Such micro-fabricated optomechanical array provides a means to realize reconfigurable long-range phonon dynamics \cite{synchronization,phonondynamics}, asymmetric and nondispersive transmission \cite{soliton,nonreciprocity2}, and dynamical phase transitions \cite{dynamical-phase-transition}.

The interaction between cavity field and mechanical oscillation in an optomechanical system can be well described by the Hamiltonian \cite{Law} of optomechanical coupling $\hat{H}_{I}=-g_0 (\hat{b}+\hat{b}^\dag)\hat{a}^\dag \hat{a}$ (we set $\hbar=1$ here), where $g_0$ is the vacuum optomechanical coupling strength, and $\hat{b}$ ($\hat{b}^\dag$) and $\hat{a}$ ($\hat{a}^\dag$) are the annihilation (creation) operator of the mechanical mode and the cavity field, respectively. In the optomechanical array, the photon hopping between neighboring optomechanical systems can be described by the Hamiltonian $\frac{J}{2}\sum_j { {\left( {{\hat{a}_j}\hat{a}_{j + 1}^\dag  + \mathrm{H.c.}} \right)} }$, where $J$ is the coupling strength of photon hopping, H.c. is the Hermitian conjugate, and the subscript $j$ denotes that the quantity described belongs to the $j$-th optomechanical system. Based on the Hamiltonian, a group of nonlinear equations can be obtained to describe the evolution of the cavity fields and the properties of the mechanical oscillation:
\begin{gather}
{\dot \alpha}_j=-(\text{i}\varpi + \frac{\kappa}{2})\alpha_j + \text{i}g_0\alpha_j \left(\beta_j + \beta_j^*\right)
-\text{i}\frac{J}{2}\left(\alpha_{j+1} + \alpha_{j-1}\right), \nonumber\\
{\dot \beta}_j=-(\text{i}\Omega_m + \Gamma_m)\beta_j + \text{i}g_0 {\left|\alpha_j\right|^2},\label{eq:0}
\end{gather}
where $\alpha_j(t)\equiv \langle \hat{a}_j(t)\rangle$ and $\beta_j(t)\equiv \langle \hat{b}_j(t)\rangle$ describe the average values of the annihilation operator of cavity fields and mechanical modes, respectively. $\varpi$ is the cavity resonance frequency, $\kappa$ ($\Gamma_m$) is the loss rate of the cavity fields (mechanical modes), $\Omega_m$ is the angular frequency of the mechanical resonance, and the quantum correlations and noises have been safely ignored in the concerned weak-coupling regime, i.e., $g_0/\kappa\ll 1$ \cite{rev}.

To describe the results more clearly, we introduce the coordinate that the $j$-th optomechanical system is located at $x_j=-1+(j-1)\Delta x$ with $\Delta x=2/(N-1)$ the distance between any two neighboring optomechanical system. Here we are interested in the continuum limit that $N$ is large enough. In this case we can use fields $\alpha(x, t)$ and $\beta(x, t)$ defined on a continuous spatial domain to describe the sets of cavity fields $\{\alpha_j\}$ and the motion of mechanical oscillators $\{\beta_j\}$. The evolution of the fields $\alpha(x, t)$ and $\beta(x, t)$ obeys a group of nonlinear partial differential equations:
\begin{gather}
-\text{i}\frac{\partial\alpha}{\partial t} = \text{i}\frac{\kappa}{2}\alpha + {g_0}\alpha \left({\beta + {\beta^*}}\right) - \frac{J}
{2}\Delta{x^2}\frac{\partial^2\alpha}{\partial{x^2}}, \nonumber\\
\frac{\partial\beta}{\partial t} = - (\text{i}{\Omega_m} + {\Gamma_m})\beta + \text{i}{g_0}{\left|\alpha \right|^2},\label{eq:1}
\end{gather}
where $({\alpha _{j + 1}} - {\alpha _j}) - ({\alpha _j} - {\alpha _{j - 1}}) \approx \Delta {x^2}  {\partial ^2}\alpha/ \partial {x^2}$ is used. This group of nonlinear equations give a continuum description of both photonic and phononic dynamics.

We seek for analytical solutions of equations (\ref{eq:1}) using the following ansatz:
\begin{gather}
\alpha (\xi ,\tau ) = \frac{{{\theta _0}{\exp({{\text{i}}\theta _0^2\tau })}\{\Xi (\xi ,\tau ){\exp[{{\text{i}}\phi (\tau )}}]\ - 1\}}}
{{{x_d}{{(\gamma /|J|)}^{1/2}}}},\nonumber\\
\beta (\xi ,\tau ) = \wp(\xi ) \exp \left[ { - \frac{{x_d^2({\text{i}}{\Omega _m} + {\Gamma _m})}}
{{|J|}}\tau } \right] + \frac{{{g_0}|\alpha (\xi ,\tau ){|^2}}}
{{{\Omega _m} - {\text{i}}{\Gamma _m}}},\nonumber
\end{gather}
where $\xi={x}/({x_d}\Delta x)$ and $\tau = |J|t/x_d^2$ are normalized space and time coordinate, respectively, $\wp (\xi ) = \beta (\xi ,0) - {\text{i}}{g_0}|\alpha (\xi ,0){|^2}/({\text{i}}{\Omega _m} + {\Gamma _m})$, $\gamma = 2g_0^2 \Omega_m/(\Gamma_m^2 + \Omega_m^2)$, and $\theta _0$ and $x_d$ are constants that describe the characteristic scale of the spatiotemporal structure. The function $\Xi (\xi ,\tau )$ can be seen as the amplitude modulation upon a plane-wave. Substitution of the ansatz into Eqs. (\ref{eq:1}) leads to two equations corresponding to the real and imaginary parts, respectively, and finally results in the following equations for $\Xi (\xi ,\tau )$:
\begin{gather}
\frac{{\partial \Xi (\xi ,\tau )}}{{\partial \tau }} - \theta _0^2\sin 2\phi \Xi (\xi ,\tau ) + \theta _0^2\sin \phi {\Xi ^2}(\xi ,\tau ) = 0, \label{eq:g1}\\
- \frac{1}{2}\frac{{{\partial ^2}\Xi (\xi ,\tau )}}{{\partial {\xi ^2}}} + \left[ {\frac{{{\text{d}}\phi }}{{{\text{d}}\tau }} - 2\theta _0^2{{\cos }^2}\phi } \right]\Xi (\xi ,\tau )\qquad\qquad\nonumber\\
\qquad + 3\theta _0^2\cos \phi {\Xi ^2}(\xi ,\tau ) - \theta _0^2 {\Xi ^3}(\xi ,\tau ) = 0.\label{eq:g2}
\end{gather}
We note that, to obtain these equations, the slowly varying approximation ${{{\text{d}}}|\alpha (x,t)|/{\text{d}}{t}} \ll \Omega _m |\alpha (x,t)|$ and the weak-decay approximation $\kappa x_{\text{d}}^2/|J| \ll 1$ are used.

Equation (\ref{eq:g1}) is a Riccati-like equation and admits the exact solution
\begin{equation}
\Xi (\xi ,\tau )=-\frac{{E(\tau )}}{{\cos ({r_0}\xi ) - F(\tau )}},\label{eq:solution1}
\end{equation}
where $\ln E(\tau ) = \theta _0^2\int {\sin 2\phi {\text{d}}\tau }$, $F(\tau ) = \theta _0^2\int {E(\tau )\sin \phi {\text{d}}\tau}$, and $r_0$ is a constant which describes the modulation frequency. The second order differentiation of $\Xi (\xi ,\tau )$ is
\begin{equation}
\frac{{{\partial ^2}\Xi (\xi ,\tau )}}{{\partial {\xi ^2}}} = \frac{{2E(\tau )r_0^2{{\sin }^2}({r_0}\xi )}}{{{{\left[ {F(\tau ) - \cos ({r_0}\xi )} \right]}^3}}} - \frac{{E(\tau )r_0^2{{\cos }}({r_0}\xi )}}{{{{\left[ {F(\tau ) - G(\xi )} \right]}^2}}},
\end{equation}
which, after some calculations, leads to the following nonlinear oscillator equation:
\begin{gather}
\frac{{{\partial ^2}\Xi (\xi ,\tau )}}{{\partial {\xi ^2}}} =  - r_0^2\Xi (\xi ,\tau ) + 3r_0^2\frac{{F(\tau )}}
{{E(\tau )}}{\Xi ^2}(\xi ,\tau ) \nonumber\\
+ 2r_0^2\frac{{1 - {F^2}(\tau )}}{{{E^2}(\tau )}}{\Xi ^3}(\xi ,\tau ).\label{eq:solution2}
\end{gather}
Comparison between equations (\ref{eq:solution2}) and (\ref{eq:g2}) gives the explicit expression of functions $E(\tau )$ and $F(\tau )$, which consequently leads to the final solution:
\begin{equation}
\alpha (\xi ,\tau ) = \frac{{{\theta _0}{e^{{\rm{i}}\theta _0^2\tau }}}}{{{x_d}{\sqrt{\gamma /|J|}}}}\left( {\frac{{2{\iota ^2}\cosh (\vartheta \tau ) - {\rm{i}}2\iota \vartheta \sinh (\vartheta \tau )}}{{\cosh (\vartheta \tau ) - \sqrt {1 - {\iota ^2}} \cos ({r_0}\xi )}} - 1} \right),\nonumber
\end{equation}
where $\iota  = {r_0}/2{\theta _0}$ and $\vartheta = - 2\theta _0^2\iota \sqrt{1 - {\iota ^2}}$. This solution is exactly the same as the famous Akhmediev breather solution \cite{Akhmediev} which describes modulation instability:
\begin{equation}\label{eq:abs}
q = \frac{{{r_1}\cosh ({r_2}\tau ) + \sqrt {2s} \cos ({r_3}\xi ) + \mathrm{i}{r_2}\sinh ({r_2}\tau )}}
{{\sqrt {2s} \cos ({r_3}\xi ) - \cosh ({r_2}\tau )}}{e^{\mathrm{i}\tau }},
\end{equation}
where ${r_1} = 1 - 4s$, ${r_2} = \sqrt {8s(1 - 2s)}$, and ${r_3} = 2\sqrt {1 - 2s}$ with the parameter $s \in (0, 0.5)$ is the modulation parameter of the Akhmediev breather solution and governs the global behavior of the Akhmediev breather.

\begin{figure}[ht]
\centering
\includegraphics[width=0.38\textwidth]{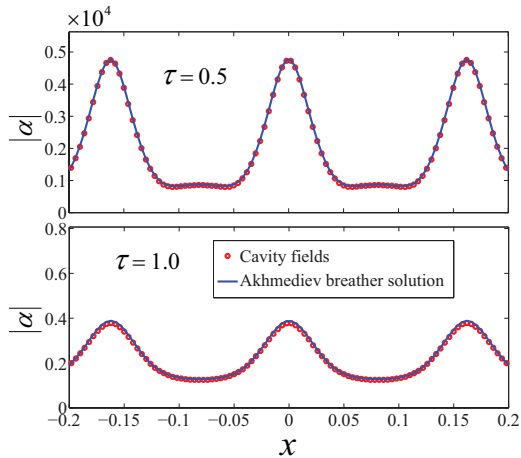}
\caption{\label{fig:2} Comparison between the cavity field distribution $\alpha(x,t)$ and analytic Akhmediev breather solution with the same parameters as Fig. \ref{fig:1}(b).}
\end{figure}

This class solution of Akhmediev breather in a micro-fabricated optomechanical array is essentially nonperturbative, and can not be obtained through perturbative methods used previously. The stability of the optomechanical Akhmediev breather can be examined by the Lie transformation \cite{Lie} or the standard spectrum theory. On the other hand, evolutionary cavity fields as well as mechanical oscillations can be obtained by solving equations (\ref{eq:0}) numerically, which confirm the Akhmediev breather behavior [shown in Fig. \ref{fig:1} (b)] in a broad parameter range.

Quantificationally, evolutionary cavity field distribution $\alpha(x,t)$ [obtained by solving equations (\ref{eq:0}) via Ronge-Kutta method] shows an excellent agreement with the analytic Akhmediev breather solution [Fig. \ref{fig:2}(a) and (b)]. The parameters of the optomechanical array \cite{parameter} used in these numerical calculations are $\Omega_m/2\pi=9.5$ GHz, $\kappa/2\pi=3.8$ MHz, $g_0/2\pi=292$ kHz, $J/2\pi=-95$ GHz, $\Gamma_m/2\pi=19$ kHz, placing it well into the experimentally relevant weak-optomechanical coupling regime $g_0/\kappa\ll 1$ which justifies the sufficiently of the classical approximation. The initial values of $\alpha_j$ is discretely chosen as the Akhmediev breather solution at time $t=0$, and the initial distribution of the mechanical oscillators is chosen to be $\beta_j= 0$ for convenience.

\begin{figure}[ht]
\centering
\includegraphics[width=0.38\textwidth]{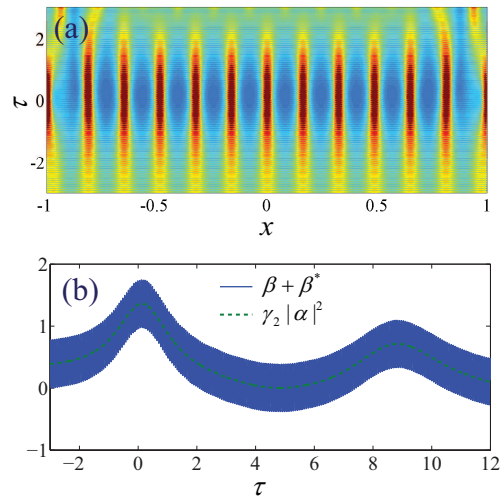}
\caption{\label{fig:3} (a) Temporal evolution of the mechanical distribution $|\beta(x,t)|$ in an optomechanical array; (b) Comparison between evolution of the cavity field $\alpha(0,t)$ and mechanical oscillation $\beta(0,t)+\beta^*(0,t)$ with $\gamma_2=\gamma/g_0$.}
\end{figure}

Temporal evolution of the mechanical distribution $|\beta_j|$ (similar for the phonon number $|\beta_j|^2$) in the micro-fabricated optomechanical array also exhibits the Akhmediev breather characteristics [shown in Fig. \ref{fig:3}(a)] where mechanical motions (described by $\beta+\beta^*$) fits in well with the time varying radiation pressure [shown in Fig. \ref{fig:3}(b)]. The phononic Akhmediev breather is also quite robust which makes things quite interesting that both photonic and phononic Akhmediev breather can be created in the optomechanical array. The experimental platform shown in Ref. \cite{crystals2} seems closest to the requirements for implementing our proposal, and tapered and dimpled optical fibers (Tapers) can be used for readout, control and signal processing \cite{crystals,crystals2,crystals3}.

We note that our result is still valid even when the dissipation (or photon hopping) strongly exceeds the optomechanical interaction. From the conventional view based on perturbative methods, there is hardly any physics that is mainly triggered by the interaction in this case. Thus our analysis based on non-perturbative approach is quite non-trivial, and may offer a more complete view of the optomechanical interaction, especially for the investigation of optomechanical nonlinearity arising from collective motions, which is entirely different from previous works \cite{hsg1,hsg2,ma}.

Optomechanical Akhmediev breather offers the ability to control both photonic and phononic localization, which is desirable for realizing oscillator networks \cite{rev3} in which the phononic behaviors can be well addressed through the photonic flow. With the improvement of nano-fabrication techniques, optomechanical arrays have the potential to be integrated with various on-chip solid-state construction, including quantum dot and superconducting quantum circuits. The combined properties of both photonic and phononic Akhmediev breather and the chip-scale size make this setup ideally suited for realizing hybrid interfaces.

These results also introduce a strong link between cavity optomechanics, hydrodynamics, and nonlinear science. Especially, the Akhmediev breather behavior is quite remarkable: it closely related to freak waves in hydrodynamics \cite{MI} and the famous Fermi-Pasta-Ulam recurrence phenomenon \cite{Fermi,Fermi2} in statistic mechanics. It is quite interesting if these remarkable phenomena can be recreated and studied in a micro-fabricated optomechanical array, and further analysis and other practical considerations are required to move in that direction.

In conclusion, we have demonstrated both numerically and analytically the emergence of novel coherent localized behaviors, optomechanical Akhmediev breathers, in a micro-fabricated optomechanical array. The optomechanical Akhmediev breather consists of a localized nonlinear structure and exhibits typical growth-return cycle characteristic. This phenomenon originates from optomechanical nonlinearity in the non-perturbative regime, and may find application in on-chip manipulation of photonic or phononic information processing.


\begin{acknowledgments}
The work was supported by the National Science Foundation of China (Grant Nos. 11774113, 11405061) and the National Key Research and Development Program of China (Grant No. 2016YFA0301203).
\end{acknowledgments}

\end{document}